\documentclass{llncs}
\usepackage{llncsdoc}
\usepackage{amssymb}
\usepackage{amsmath}
\usepackage[final]{listings}
\usepackage[normalem]{ulem}
\usepackage[all]{xy}

\newcommand{\mathset}[1]{\left\{#1\right\}}
\newcommand{\mathpset}[2]{\left\{#1 \mid #2 \right\}}

\newtheorem{algorithm}{Algorithm}
\DeclareMathOperator\closure{cl}

\begin{document}

\title{Integrating Space, Time, Version and Scale Using Alexandrov Topologies}
\author{Norbert Paul
       \and Patrick Erik Bradley
       \and Martin Breunig
       }

\institute{Karlsruhe Institute of Technology, 76128 Karlsruhe, Germany}

\maketitle

\begin{abstract}
As a contribution to higher dimensional spatial data modelling this article 
introduces a novel approach to spatial database design. 
Instead of extending the canonical Solid-Face-Edge-Vertex 
schema of topological data, 
these classes are replaced altogether by a common type \texttt{SpatialEntity}, 
and the individual ``bounded-by'' relations between two consecutive 
classes are replaced by one 
separate binary relation \texttt{BoundedBy} on \texttt{SpatialEntity}. 
That relation defines a so-called \emph{Alexandrov topology} 
on \texttt{SpatialEntity} and thus exposes the fundamental 
mathematical principles of spatial data design. 
This has important consequences:
First, a formal definition of topological ``dimension'' for 
spatial data can be given.
Second, every topology for data of arbitrary dimension has such a simple 
representation. 
Also, version histories have a canonical Alexandrov topology,
and generalisations can be consistently modelled by the new consistency rule 
\emph{continuous functions} between LoDs, 
and \emph{monotonicity} enables accelerated path queries. 
The result is a relational database schema for spatial data of 
dimension 6 and more which seamlessly integrates 4D space-time, levels 
of details and version history.
Topological constructions enable queries across these different aspects.


\end{abstract}

\section{Introduction}


2D and 3D spatial models are well established for spatial data modelling, 
and there exist 
standards  like CityGML in geo-spatial modelling and  
IFC for architectural models.  
Currently there is active research on spatio-temporal queries \cite{SG2009}
as well as 4D spatio-temporal modelling \cite{Anh:4D}, and also on 
considering other aspects 
like scale \cite{Oosterom:5D} 
as additional dimensions 
of spatial data. 
Also integrating the versioning graph,  by itself  1D, 
 pushes up the dimension upper bound. 

Besides that, research on $n$D modelling provides generic spatial data models without a fixed 
dimension upper bound \cite{Lienhardt:GMaps} and often gives a formal definition of 
``topological dimension'' of spatial data. Topology has its own sub-discipline called ``dimension theory'' 
\cite{Engelking:DimensionTheory} where the possible definitions of such ``topological dimension'' are 
investigated. 
Among these, the \emph{Krull dimension} \cite[p.\ 5]{Hartshorne:AG} 
is particularly 
applicable for topological data and  is proposed here as a standard 
definition of spatial data dimension. 
Throughout this article ``dimension'' of spatial data is the 
Krull dimension of the topological space 
established by the data entities. 

Challenged by a scientist's request to show that there are applications for 
data of dimension beyond 4, 
and inspired by \textsc{van Oosterom}'s ideas \cite{Oosterom:VarioScale}, 
this article demonstrates the mathematical foundations of generic $n$D spatial 
modelling and its use for combining 3D spatial data, time, scale, 
and versioning into an 
integrated 6D+ model. 
In particular, it will be shown that sensible integration of scale increases 
the dimension of spatial data by more than one, hence the ``+'' in 6D+.  
Even more, further increasing the dimension can \emph{decrease} 
the complexity of the data model, 
which renders 
statements like 
``$n$D modelling leads to a combinatorial explosion of complexity'' 
wrong. 

The reader is assumed familiar with 
basic concepts of mathematical topology. 
To facilitate the lecture, references are given to locations in 
textbooks 
explaining the applied concepts in more detail. 
As the main intent is to expose the mathematics of $n$D modelling, 
the relational model, 
formulated by {\sc Codd} \cite{Codd:RMV2}, is used. 
Its sound mathematical basis 
alleviates its topologising in an elegant way \cite{BradPaul:RelTop}, and 
the principles described here are also applicable to object oriented modelling.
They  also serve as a basis for a generic topological database model as an extension 
of the relational model: instead of tables, 
there are spaces on which queries operate 
and which they return as a result. 
The first author of this article 
has implemented an experimental prototype of such a topological relational 
database model which runs on \url{http://pavel.gik.kit.edu}.



\section{Dimension}\label{sec:dim}

In 3D spatial data usually four kinds of 
topological entities are represented in computer storage: 
\emph{Vertices} are zero-dimensional discrete points in $\mathbb{R}^3$. 
\emph{Edges} are one-dimensional manifolds which are the interior of paths 
starting at one vertex and ending at another vertex. 
In general, an edge is \emph{bounded} by two vertices. 
\emph{Faces} are two-dimensional manifolds 
enclosed by at least one 
loop of edges. Some models allow to specify more loops where 
one loop is the outer boundary and each additional loop is a hole boundary. 
Topologically, there is no difference between an ``outer'' boundary loop and a ``hole'' boundary 
loop because every hole can be made an outer boundary by stretching it wide enough 
and then flipping the face over.
To wrap this up, a face is bounded by edges. 
\emph{Solids} are three-dimensional manifolds enclosed by a set of faces 
which constitute a cavity within which the solid resides. 
Such cavity is often called a \emph{shell} and, 
again, a solid is bounded by faces. 
Most 3D models establish a ``chain'' of four classes 
with two consecutive classes 
connected by a ``bounded-by'' association 
(cf.\ \cite{Zlatanova:Classical} for an overview). 
Note that the chain length equals the model dimension. 


According to 3D spatial models, time can be modelled by 
the real line $\mathbb{R}$. 
Each moment in time can be represented by a real number, 
thus 
resembling
a vertex in $\mathbb{R}^3$. A time span is an open interval $(t_1,t_2)$ 
bounded by a 
starting point $t_1$ and an ending point $t_2$, 
thus 
resembling an edge. 
This gives two classes of temporal entities: \emph{Moment} 
and \emph{Timespan} where each 
one-dimensional time\-span is bounded by two zero-dimensional moments. 
According to spatial data this 
``bounded by'' association 
can be considered a special case 
of an association chain of length 1 which, again, 
corresponds with the dimension. 


When data is changed over time several versions of the data exist. 
In versioning software two consecutive versions along a version history are commonly 
connected by an edge. 
Versions can also fork and merge and so a version history is a directed acyclic graph 
(DAG) with two classes, a \emph{Version} and a \emph{Transition} and each transition is 
bounded by an initial and a terminal version. 
So we could say that the ``dimension'' of a version history is one: there is only one association 
from \emph{Transition} to \emph{Version}. 
We will later take an alternative view on the version history which 
results in a different dimension. 


Spatial data is often organised hierarchically at different levels of ``scale'' or ``detail'' (LoD). 
For smooth transitions between LoDs it is often proposed to interpolate 
between consecutive LoDs by continuous morphing \cite{Oosterom:VarioScale}. 
Then the space between two consecutive LoDs is considered an edge 
bounded by two LoDs. So, we have a one-dimensional space. 


Thus we have four types of spaces which we might call ``elementary spaces'': 
The ``spatial'' space which is the Euclidean 3D, the one-dimensional Euclidean temporal space, 
a one-dimensional version space, and another one-dimensional scale space which is essentially 
a linear graph $\cdots\to\bullet\to\bullet\to\cdots$. 


Now a combination of these elementary spaces can be used to get higher-dimensional 
``combined spaces''. 
In such a combined space a 3D \emph{Solid} $s$ and a 1D \emph{Timespan} $ts$ 
can be combined to a 4D pair $(s,ts)$ which represents a 4D entity in space-time: the trajectory of 
$s$ during $ts$. 
But for a zero-dimensional \emph{Moment} in time $t$, the pair 
$(s,t)$ is a 3D element in space-time representing 
that solid $s$ at the moment $t$. When $ts$ is 
bounded by $t$ then the 4D-entity $(s,ts)$ is bounded by a 3D-entity $(s,t)$ and for each lower-dimensional 
$n$D element $x$ in the bounded-by association chain of $s$ there is a pair $(x,t)$ in a corresponding 
association chain of $(s,t)$. This means that in the 4D model the length of the association chains has 
increased by 1. 
Within that model each pair $(a,b)$ satisfies the dimension formula $\dim(a,b) = \dim a + \dim b$. 


As every database management system (DBMS) can only model finite sets we need a definition of 
``dimension'' for finite spaces. 
As seen above, our $n$D spaces consist of entities, possibly distributed over several 
classes, and a bounded-by relation of chain length $n$. 
Hence ``dimension'' of spatial data is the maximal length of a chain of entities such that 
each is bounded by a consecutive element. 
We call this the \emph{combinatorial dimension} of spatial data. 
The note \cite{BradPaul-Krull} proves that this combinatorial dimension is equivalent to the 
topological Krull dimension \cite[p.\ 5]{Hartshorne:AG}. 
Note that even a simple sequence $(n,n-1,\dotsc,0)$ can be considered a set 
$\mathset{n,n-1,\dotsc,0}$ with ``association'' $a\,S\,b \Leftrightarrow  a - 1= b$ which then
has a combinatorial dimension of $n$ and obviously does 
not suffer from any ``combinatorial explosion of complexity'' whatsoever. 
We 
can even decrease the complexity of 
spatial data by increasing its dimension.


\section{Spatial Dimension and Consistency}\label{sec:spadim}

Usually, in a DBMS consistency rules are design tools, 
and it lies at the discretion of the user to make use of them. 
However, in spatial data modelling, 
``topological consistency'' is often mentioned, but the rules to 
tell ``consistent'' from ``inconsistent'' spaces vary in the literature. 
We see two reasons: 
First, as mentioned above, the user should be entitled to decide which spatial 
data he considers ``consistent'' and which he does not. 
Second, the term ``topological (in)consistency'' is unknown in mathematics.  
Topology provides a rich set of well-defined topological \emph{properties} 
which may or may not be used for a particular 
application as a consistency rule. 
We will discuss here some ``consistency'' for $n$D spatial 
entities and present applications where these 
rules do not apply. 

A \emph{vertex} is usually composed of an id, the  coordinates, 
and additional attributes. 
The consistency rule is that a vertex is not bounded by another object. 
However, this property can be used to \emph{define} vertex: 
In a topological space a \emph{vertex} is an element that 
is not bounded by another element. 
A practical application is the room connectivity graph of a building: 
In this graph the rooms are the 0D vertices, and the doors are their connecting 
1D edges. Each door is then ``bounded by'' its two rooms. 
This possibility of ``shifting down'' the dimension and inverting it by ``flipping'' the 
boundary relation is a characteristic of the topological spaces occurring in spatial 
data models. That ``flipping'' is intimately related to the Poincar\'e duality 
with which it should not be confounded. 

Some spatial data models define faces by a cycle of vertices around that face. 
This gives implicit edges between two consecutive vertices in the cycle. 
But we will only consider models with explicitly given edges. 
Then an edge usually has two references to two boundary vertices. 
When we weaken that rule of exactly two vertices we 
can \emph{define} ``edge'' as a spatial entity 
with non-empty boundary that consists only of vertices. 
This  allows selection results where the selection predicate only applies 
to, say, two edges but only one common connecting vertex. 
It would then be unwise to forget the connectivity 
information in the selection result only because 
it fails to satisfy the 
global
constraints.  
Moreover, if an edge is allowed to have more than one vertex, 
then the data model allows hypergraphs, 
an abstract topological structure often used in practice. 
Additionally, an edge with four vertices might be considered 
an edge with a hole that occurs e.g.\ in intersections with non-convex faces. 

The classical consistency rule for faces is that the boundary edges must form loops. 
Mathematically, this means that a boundary must form a cycle, 
which is the fundamental property of a \emph{chain complex} from algebraic topology 
\cite{Hatcher:AT}. 
However, it makes sense in general to permit query results violating consistency rules 
that have been set up for the underlying query input spaces. 

Another consistency rule is hard-coded into the classical sequential class schema: The 
only possible association between a face and one of its boundary vertices is indirectly via an 
intermediate edge. The schema does not allow a direct face-vertex association that ``bypasses'' the 
edge class. However, a vertex within a face may be considered, say, a collapsed interior face, 
e.g.\ a city in a region at a lower LoD. 
If we now specify a superclass \emph{SpatialEntity} of 
Solid, Face, Edge, and Vertex 
and replace the associations between two consecutive 
classes by one association of that class to itself, 
that rule can be made optional, too. 

Making the above consistency constraints optional immediately leads to a 
directed acyclic graph (DAG) of spatial entities. 
A DAG defines a partial ordering on its elements, and since 1937 it is well-known that 
partial orderings are essentially the same as the so-called 
\emph{$T_0$ Alexandrov topologies} \cite{Alex:DS}, or, 
to put it short, \emph{spatial data models are topological spaces}. 
We will now provide an initial relational database schema for 3D spatial data based on this observation:
\begin{eqnarray}
 &X(\underline{\textit{id}},\textit{attributes}),
 R(\underline{\textit{ida}},\underline{\textit{idb}}),\quad
 (\textit{ida})\stackrel{\textit{FK}}{\longrightarrow} X,\;
 (\textit{idb})\stackrel{\textit{FK}}{\longrightarrow} X \label{eqn:dtopmodel}
\\\nonumber
&\mathit{Vertex}(\underline{\textit{vid}},
                           \textit{x}:\mathbb{R},
                           \textit{y}:\mathbb{R},
                           \mathit{z}:\mathbb{R}),\quad
 (\textit{vid})\stackrel{\textrm{FK}}{\longrightarrow} X \enspace .
\end{eqnarray}
Table $X$ contains the spatial entities, 
and  table $R$ specifies the ``bounded-by'' relation. 
Primary key attributes are underlined. 
Attribute \textit{attributes} is a placeholder for 
the set of ``semantic'', i.e.\ non-spatial attributes. 
The only consistency rule 
here is that $R$ should be acyclic. 
Although every relation $R$ defines an Alexandrov topology $\mathcal{T}(R)$ of the space 
\begin{eqnarray*}
(X,\mathcal{T}(R)),\quad
&\mathcal{T}(R) := 
\mathpset{A\subseteq X}{\forall(a,b)\in R: b\in A \Rightarrow a \in A}
\end{eqnarray*} 
that space is $T_0$-separable iff $R$ is acyclic \cite[Ex.\ 133]{Adamson:TW}. 
$T_0$-separability is a sensible consistency rule for spatial data, and will be maintained here. 
Note, however, that non-separable spaces also  are valid topological spaces, 
and so from the mathematical viewpoint acyclicity is an optional consistency rule, too. 
For each topology-defining relation $R$ on $X$ we assume the existence of a view on its pre-order 
$R^{*}$ which is the transitive and reflexive closure of  $R$:
\begin{lstlisting}
   create view poR as 
   with recursive Rst(ida,idb) as (
      select id as ida, id as idb from X
      union 
      select Rst.ida, R.idb 
      from Rst join R on (Rst.idb = R.ida))
   select * from Rst;
\end{lstlisting} 
This schema allows spatial data of arbitrary dimension, but for the moment we assume 
that $R$ chains have length $\le 3$, so the topological dimension matches the ``geometric dimension'', 
i.e.\ the number of coordinate attributes 
of $\mathit{Vertex}$. 
Also, the \textit{id} of all vertices should occur in table $\mathit{Vertex}$. 
For obvious reasons, we  call a pair $(X,R)$ 
consisting of a set $X$ and a binary relation $R$ on $X$ a 
\emph{topological data type}, or sometimes simply \emph{space}.


\section{Temporal Dimension}\label{sec:tempdim}

\begin{figure}[t]
$$
\newcommand{\ds}{\displaystyle}
\xymatrix@R=0pt{
  &           & w_l \ar@{}[r]|{\ds{I}} & w_r\\
  &\ar@{-}[r] & \bullet \ar@{=}[r] &\bullet \ar@{-}[rr]&&&\\
\\
\\  &           & (w_l,t_0)\  \ar@{}[r]|{\ds{(I,t_0)}}  & \ (w_r,t_0)
\\  &\ar@{-}[r] & \bullet \ar@{=}[r] 
                          \ar@{=}[dddd]_{\ds{(w_1,s_{0,1})}}
                          \ar@{}[dddddddd]_{\ds{(w_1,t_1)}}
                                   &\bullet \ar@{-}[rr] \ar@{=}[dddd]^{\ds(w_r,s_{0,1})}
                                                      && t_0
\\  &           &                  &                  &&
\\  &           & \ar@{}[r]|{\ds{(I,s_{0,1})}}
                                   &                  && s_{0,1}             
\\  &           &                  &                  &*{(w_{rr},t_1)}&
\\  &           & \bullet 
                  \ar@{=}[r]_{\ds{(I,t_1)}}
                  \ar@{}[rr]_{\ds{(w_r,t_1)}}
                  \ar@{=}[dddd]_{\ds{(w_1,s_{1,2})}}
                                   & \bullet 
                                     \ar@{=}[r]^{\ds{(X,t_1)}}
                                     \ar@{}[dddd]|{\ds{(J,s_{1,2})}}
                                                      & \bullet
                                                        \ar@{-}[r]
                                                        \ar@{=}[dddd]^{\ds{(w_{rr},s_{1,2})}}
                                                      & t_1                 
\\ &           &                   &                  &               &     
\\ &           &                   &                  &               & s_{1,2}
\\ &           &                   &                  &               &
\\ &\ar@{-}[r] & \bullet\ar@{=}[rr]&                  &\bullet\ar@{-}[r]& t_2
\\ &           & (w_l,t_2)\  \ar@{}[rr]|{\ds{(J,t_2)}} & & \ (w_{rr},t_2)
\\
\\
\\
  &           & (w_l,t_0)\  \ar@{}[r]|{\ds{(I,t_0)}}  & \ (w_r,t_0)\\
  &\ar@{-}[r] & \bullet \ar@{=}[r] \ar@{=}[dddddddd]_{\ds(w_1,s_{0,2})}
                                   &\bullet \ar@{-}[rr] \ar@{=}[dddd]^{\ds(w_r,s_{0,1})}
                                                      && t_0\\
  &           &                    &                  &&\\
  &           &                    &                  && s_{0,1}\\
  &           &                    &                  &*{(w_{rr},t_1)}&\\
  &           &    \ar@{}[r]|{\ds{(\mathit{IJ},s_{0,2})}}
                                   &\bullet \ar@{=}[r]_{\ds{(X,t_1)}}
                                                      &\bullet\ar@{-}[r]
                                                              \ar@{=}[dddd]^{\ds{(w_{rr},s_{12})}}
                                                       & t_1\\
  &           &                    &*{(w_r,t_1)}          &&\\
  &           &                    &                  && s_{1,2}\\
  &           &                    &                  &&\\
  &\ar@{-}[r] & \bullet\ar@{=}[rr] &                  &\bullet\ar@{-}[r]& t_2\\
  &           & (w_l,t_2)\  \ar@{}[rr]|{\ds{(J,t_2)}}  && \ (w_{rr},t_2)\\
}
$$
\caption{The process of a 1D house in Lineland constructed at time $t_0$, 
extended by a portion $X$ at time $t_1$, 
and demolished at time $t_2$, is modelled as a 2D space-time 
complex in two steps. The middle complex repeats each element at a 
given point or period of time, 
and the lower complex identifies some elements, the left wall $w_l$ and some 
elements of the interior $I$, 
to get fewer entities to store.}\label{fig:llhouse}
\end{figure}
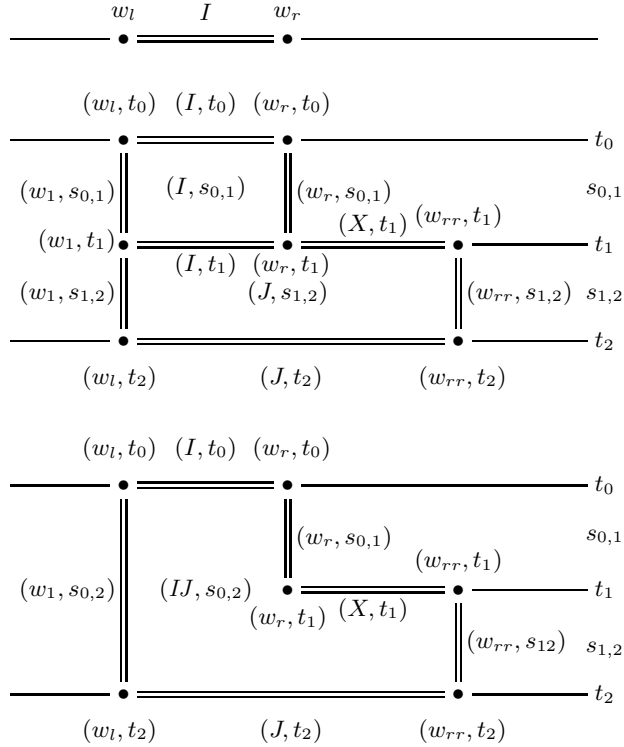

Here, we establish the 4D \emph{space-time} which models changes of  
3D space over time. 
For illustrative reasons,
a 2D space-time of a 1D ``building'' example in 
Lineland \cite[ch.\ 13]{Abbott:Flatland} changing over 1D time
will be discussed first. 
Assume in 1D Lineland 
a ``house'' with interior $I$ and boundary walls $w_l$ to 
the left and $w_r$ to the right as depicted on top of Fig.\ \ref{fig:llhouse}. 
The boundary of 
$I$ are the 
vertices $w_l$ and $w_r$. 
The house is erected at 
time $t_0$, modelled by ``tagging'' 
$I$, $w_l$ and $w_r$ with $t_0$ giving the 
pairs $(I,t_0)$, $(w_l,t_0)$, and $(w_r,t_0)$. 
The spatial ``bounded-by'' associations from $I$ to $w_l$ and 
from $I$ to $w_r$ are carried over 
as tagged space-time associations 
from $(I,t_0)$ to $(w_l,t_0)$ and from $(I,t_0)$ to $(w_r,t_0)$. 
The interior $I$, existing over a time-span $s_{0,1}$, 
 is also tagged,  giving its trajectory $(I,s_{0,1})$. 
This element is bounded in the horizontal (i.e.\ ``spatial'') 
direction by $(w_1,s_{0,1})$ because $I$ is bounded by $w_1$. 
Here the ``tag'' $s_{0,1}$ does not change in this boundary association. 
The element $(I,s_{0,1})$ is also bounded in the vertical 
(i.e.\ ``temporal'') direction by $(I,t_0)$. 
Here the boundary association fixes element $I$ 
and is taken from the boundary association between $s_{0,1}$ and $t_0$. 

The space at time point $t_1$ models the before-after change 
derived by a 1D overlay of the spatial model of $I$ before,  
and the spatial model 
of $J$ after the change. Each overlay entity has a reference to 
its corresponding entity 
of the input spaces. Mathematically, these are two topologically 
continuous partial functions 
$p$ (like ``past'') and $f$ (like ``future'') from the overlay space 
back to the two input spaces. 
Such functions are called \emph{attaching maps}. Now we can topologically paste 
\cite[Ch.\ 3 \S 7]{Jaenich:Topology} the overlay onto 
the ``past'' entities by specifying that a tagged image element 
$(I,s_{0,1})$ is bounded by an 
element $(x,t_1)$ if the reference $p(x)=I$ holds. We can also do that 
attachment onto the ``future'' 
by specifying that $(w_1,s_{0,1})$ is bounded by $(w_1,t_1)$ 
because $p(w_1)=w_1$. 
Interestingly $(w_r,t_1)$ is a  boundary element of the past wall trajectory 
$(w_r,s_{0,1})$ because of 
$p(w_r)=w_r$, but it is a boundary element of the future interior trajectory 
$(J,s_{1,2})$ because of $f(w_r)=J$. 

This step is only an intermediate formal step which would 
create a lot of redundant data when 
implemented explicitly. To avoid redundancy a sequence of a 
tagged spatial entity 
not changing
over time can be collapsed into one. Additionally, the user may specify 
further entities that, though having changed over time, 
are considered ``identical'' before and after the change. 
In our example this identification is carried out on 
$(w_1,s_{0,1})$, $(w_1,t_1)$, and $(w_1,s_{1,2})$, and 
it collapses $(I,s_{0,1})$, $(I,t_1)$,
and $(J,s_{1,2})$ to $(\mathit{IJ},s_{01,2})$. 
Mathematically, this gives a topological ``quotient space'' 
\cite[Ch.\ 3]{Jaenich:Topology} 
depicted in the lower part of Fig.\ \ref{fig:llhouse}. 

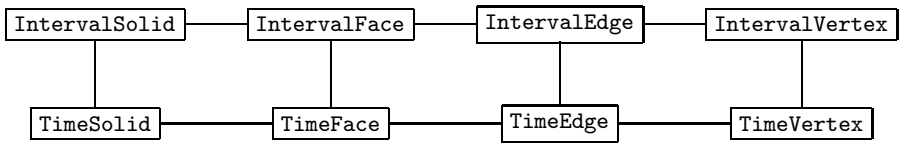
\begin{figure*}[h]
$$
  \xymatrix{
    *+[F]{\tt IntervalSolid}
    \ar@{-}[r]\ar@{-}[d] & *+[F]{\tt IntervalFace}
                           \ar@{-}[r]\ar@{-}[d] & *+[F]{\tt IntervalEdge}
                                                  \ar@{-}[r]\ar@{-}[d] & *+[F]{\tt IntervalVertex }
                                                                         \ar@{-}[d] \\
    *+[F]{\tt TimeSolid}
    \ar@{-}[r]           & *+[F]{\tt TimeFace}
                           \ar@{-}[r]           & *+[F]{\tt TimeEdge}
                                                  \ar@{-}[r]            & *+[F]{\tt TimeVertex}
    }
$$
  \caption{Directly combining the classical sequential space model 
with the same classical approach  
for a temporal model creates eight classes for time-space entities.}
  \label{fig:stgrid}
\end{figure*}

\begin{figure*}[t]
$$
  \xymatrix{
    *+[F]{\tt SpaceTimeEntity}\ar@{-}`/0pt[r]^(.65){*}`[rd]
                                     `[d]_{\displaystyle\tt \triangleleft\ BoundedBy}
                                      []^(.65){*}_(.65){\tt boundary} &&
\\                            &
  }
$$
  \caption{The one space-time class and bounded-by association 
obtained by combining the spatial and temporal models in a non-classical way.}
  \label{fig:stent}
\end{figure*}
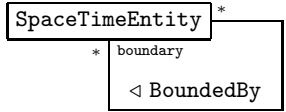

The same construction applies 
to 3D spatial models to get a 4D space-time which 
is simply another topological space. 
However, when the 3D model is represented by a chain of three 
associations, and when the temporal 
model also has that classical layout, we get a ``grid'' of eight different 
classes of pairs of spatial-temporal entities 
(cf.\  {Fig.\ \ref{fig:stgrid}}). This results in ten different 
space-time-boundary associations, represented by lines between classes. 
On the other hand, if we had only 
two classes \texttt{SpatialEntity} 
and \texttt{TemporalEntity}, each with a relation \texttt{BoundedBy}, 
then there would  be 
only one spatio-temporal class where each entity consists of a pair 
(\texttt{spatial},\texttt{temporal}), 
and only one boundary relation associating
each pair \texttt{(spatial,temporal)} with all 
\texttt{($\mathit{ds}$,temporal)}
for every boundary element $\mathit{ds}$ of \texttt{spatial} and, 
dually, with the pairs \texttt{(spatial,$\mathit{dt}$)} 
for every boundary time point $\mathit{dt}$ of \texttt{temporal}. 
This yields
the topology of the so-called 
\emph{product space} \cite[Ch.\ I \S 3]{Jaenich:Topology}. 
The class \texttt{SpaceTimeEntity} with a relation 
\texttt{BoundedBy} (cf.\ Fig.\ \ref{fig:stent}) allows
to model arbitrary space-time configurations.
When we consider $\mathit{id}$ a surrogate key for pairs of space-time-entities 
the relational schema (\ref{eqn:dtopmodel}) for 3D spatial data can be easily extended to 
cover spatio-temporal data: 
\begin{eqnarray*}
\mathit{Vertex}(\underline{\textit{vid}},x:\mathbb{R},y:\mathbb{R},z:\mathbb{R},t:\mathbb{R}) \enspace .
\end{eqnarray*}
Note the simplicity. 
Like \cite{SG2009}, it allows moving objects
 between two time points that may 
be geometrically interpolated by a query. 
A simple topological SQL-query 
for the space at a given time point $t$ with no interpolation 
is: 
\begin{lstlisting}
   create view Xt as 
   with mmX(id, tmin, tmax) as (
      select X.id, min(V.t) as tmin, max(V.t) as tmax
      from   (X join poR on (X.id = poR.ida)) 
             join Vertex V on (poR.idb = V.vid)
      group by X.id)
   select X.* from X join mmX on (X.id = mmX.id) 
   where (mmX.tmin < t and t < mmX.tmax)
      or (mmX.tmin = t and t = mmX.tmax); 
\end{lstlisting}
The sub-query $\mathit{mmX}$ first associates each element $\mathit{X.id}$ with 
its topological closure $\closure(\mathit{X.id})$ by joining it to 
$\mathit{poR}$. 
The join with $\mathit{Vertex}$ selects all vertices of 
$\closure(\mathit{X.id})$. 
The group-by clause then computes the time interval for $\mathit{X.id}$.

This query selects a sub\emph{set} $\mathit{Xt}$ of $X$.
As the pair $(X,R)$ defines a topological space $(X,\mathcal{T}(R))$, 
there should be a relation $\mathit{Rt}$ that generates the sub\emph{space} 
topology $\mathcal{T}(R)|_{\mathit{Xt}}$ 
(cf.\ \cite[Ch.\ I \S 3]{Jaenich:Topology}). 
Simply restricting $R$ to $\mathit{Xt}$ is generally wrong 
\cite{BradPaul:RelTop}. 
Naively restricting $\mathit{poR}$ to $\mathit{Xt}$ is correct but expensive. 
An optimal $\mathit{Rt}$ is achieved by passing that restriction 
to \textsc{Codd}'s \texttt{OPEN} operator which 
returns a minimal relation whose transitive closure 
is the same as that of the input relation \cite[p.\ 427]{Codd:ExRelDB}. 
Hence 
$\mathit{Rt} 
  := \mathtt{OPEN}(\mathpset{(a,b)\in\mathit{poR}}{a\neq b, 
                                                   a \in\mathit{Xt}, b\in \mathit{Xt}})
$.

Just as the set $\mathit{Xt}$ is a $\Theta$-selection of $X$, 
the space $(\mathit{Xt},\mathit{Rt})$ can be considered a topological $\Theta$-selection of 
$(X,R)$, hence a ``topologised'' basic relational query operator. 
The topology $\mathcal{T}(\mathit{Rt})$ is the minimal topology 
for which the inclusion function 
$i:\mathit{Xt}\to X$ is continuous \cite[Ex.\ 1]{Adamson:TW}. 
A relationally complete topological database query language 
is given in \cite{BradPaul:RelTop}. 
A first simple prototype implementation can be found under 
\texttt{http://pavel.gik.kit.edu/}.

It is also noteworthy that an edge in $X$, representing 
the trajectory of a vertex $v$ from $t_0$ to 
$t_1$ with $t_0 < t < t_1$, topologically becomes a vertex in the result space 
$\mathit{Xt}$ which does not contain the edge's space-time-vertices. 



\section{Versioning}\label{sec:version}

The \emph{version graph} is a DAG whose
vertices resp.\ edges correspond 
to the different versions resp.\ modifications of a 
space.
An edge $(i,j)$ indicates that version $i$ has been modified 
to obtain version $j$.
Clearly, not all versions need to be stored explicitly.
It suffices to store the initial version, 
and any  other version $v$ can  be reconstructed from 
it by applying all modifications on the paths to $v$ in the version graph.
%
It is also advisable to redunantly store more versions e.g\ the current version
but this leads to the problem of balancing redundancy avoidance 
against robustness and speed and we will not delve into this matter. 
Note that we only consider topological changes here.

The possible topological modifications of a pair $(X,R)$ are: 
First, a point can be added, and a point $x\in X$ can be removed. 
The latter forces relation $R$ 
to be modified by removing 
all pairs $(y,x)$ and $(x,z)$ with $y,z\in X$.
But if  $y,z$ are different from $x$ with $(y,x)\in R$
and $(x,z)\in R$, then the pair $(x,z)$ must be added to $R$ 
if it is not already contained in $R$.
The reason is that the modified set $X\setminus\mathset{x}$ 
should be a \emph{subspace}
of $(X,R)$,  meaning that the (possibly indirect) bounded-by association
$y\to z$ in $X$ must be retained in $X\setminus\mathset{x}$ by the induced
bounded-by relation.
Further, a pair of points (either new or not removed from $X$) 
can be added to $R$, 
and a pair can be removed from $R$. 
These are the elementary modifications, and a general modification
is a sequence of elementary modifications.

The following relation schema:
\begin{eqnarray*}
 X(\underline{\textit{id}},{\textit{version}},\textit{atts}),\;
\textit{version}\stackrel{\textrm{FK}}{\longrightarrow} \textit{VX},\;
 R(\underline{\textit{ida}},\underline{\textit{idb}}),\;
 \textit{ida}\stackrel{\textrm{FK}}{\longrightarrow} X,\;
 \textit{idb}\stackrel{\textrm{FK}}{\longrightarrow} X
\end{eqnarray*}
stores the versions of a space.
The \textit{version}-attribute 
gives the version 
in which an object or bounded-by association
appears for the first time, and \textit{VX} is the object table 
of the \emph{version space}
described below. 
This yields a single space 
containing all objects appearing in some version. 
The following schema 
stores when (i.e.\ in which versions) an object
or bounded-by association is deleted:
\begin{eqnarray*}
\textit{DelX}(\underline{\textit{id}},\textit{version}),
\;\textit{id}\stackrel{\textrm{FK}}{\longrightarrow} X,\;
\textit{DelR}(\underline{\textit{ida}},\underline{\textit{idb}},\textit{version}),\;
(\textit{ida},\textit{idb})\stackrel{\textrm{FK}}{\longrightarrow}R
\end{eqnarray*}
The version numbers and transitions are stored as:
\begin{eqnarray*}
\textit{VX}(\underline{\textit{version}}),\;
\textit{VR}(\underline{\textit{fromv}},\underline{\textit{tov}}),\;
\textit{fromv}\stackrel{\textrm{FK}}{\longrightarrow}\textit{VX},\;
\textit{tov}\stackrel{\textrm{FK}}{\longrightarrow}\textit{VX}
\end{eqnarray*} 
which is the
 \emph{version space} $V=(\textit{VX},\textit{VR})$,
 a topological data type for spaces of arbitrary 
combinatorial dimension. 
Since the version graph is a directed acyclic graph,
the consistency rule for the version space is that of a $T_0$-space.

The \emph{topological merge} of versions $(X_1,R_1),(X_2,R_2)$ which are 
each modifications of a common space $(X,R)$ 
is a space $(Y,S)$ fitting into the
diagram:
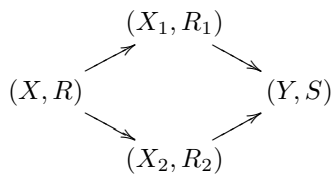
\begin{equation*}
\xymatrix@=10pt{
&(X_1,R_1)\ar[dr]\\
(X,R)\ar[ur]\ar[dr]&&(Y,S)\\
&(X_2,R_2)\ar[ur]
}
\end{equation*}
The set $Y$ consists of all points of $X$ 
occurring 
in 
$X_1$ or $X_2$, together with any additional points in either version.
A \emph{conflict} can occur if a 
point $y$ is introduced in version $X_1$,
and 
also in $X_2$. In that case, $y$ will have the same
\textit{id}-value in both versions, and the conflict 
does occur 
if some
attribute other than the \textit{version}-value does not coincide in both versions. 
We call this 
an \emph{inherent conflict}.
In the same way, the topology-defining relation $S$ is given as the set of 
all pairs from $R$ occurring in $R_1$ or $R_2$, together with all new pairs from  $R_1$ or $R_2$. 
Again, an inherent conflict occurs for pairs
$(a,b)$ appearing in both relations $R_1$ and $R_2$, if there exist
attributes other than \textit{version} and \textit{id} which are different.

If there are no inherent conflicts, then the merge $(Y,S)$ is a valid topological data type.  
However, there is a potential source for another type of conflict
which 
we call 
\emph{consistency conflict}.
This 
happens if the topological space defined by
$(Y,S)$ violates some pre-assigned 
consistency rule.
In the case of a conflict, 
a warning statement is issued, 
and it is up to the user to resolve the matter.

\begin{figure}[h]
\newcommand{\ds}{\displaystyle}
\begin{align*}
\xymatrix@C=10pt@R=.2pt{
&&&&&\stackrel{\ds h}{1}\ar[r]&\stackrel{\ds e}{2}\ar[r]&\stackrel{\ds l}{3}\ar[r]
&\stackrel{\ds p}{6}
\\
&&&&&&&&\ar[dr]\\
&&&&\ar[ur]&&&&&&&\stackrel{\ds e}{2}\ar[dr]&&\stackrel{\ds p}{6}\\
\stackrel{\ds h}{1}\ar[r]&\stackrel{\ds e}{2}\ar[r]&\stackrel{\ds l}{3}\ar[r]
&\stackrel{\ds l}{4}\ar[r]&\stackrel{\ds o}{5}
&&&&&&
\stackrel{\ds h}{1}\ar[ur]\ar[dr]&&\stackrel{\ds l}{3}\ar[ur]\ar[dr]\\
&&&&\ar[dr]&&&&&&&\stackrel{\ds a}{7}\ar[ur]&&\stackrel{\ds o}{5}\\
&&&&&&&&\ar[ur]\\
&&&&&\stackrel{\ds h}{1}\ar[r]&\stackrel{\ds a}{7}\ar[r]&\stackrel{\ds l}{3}\ar[r]
&\stackrel{\ds o}{5}
}
\end{align*}
\caption{A topological merge of texts violating the consistency rule ``linear DAG''.}\label{textmerge}
\end{figure}
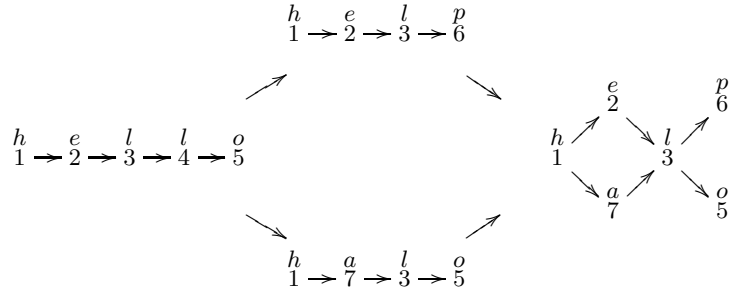

The merge of texts can be viewed as a topological merge by viewing a string
as a linear DAG $\bullet\to\bullet\to\dots\to\bullet$ whose semantic attribute takes values in an alphabet.
If the resulting space is again a linear DAG without inherent conflict, 
then the topological merge of texts is valid.
However, even without inherent conflict 
one can obtain topological spaces 
not representing text. E.g.\ 
the  topological merge of texts 
in Fig.\ \ref{textmerge} exhibits 
this consistency conflict.

Observe that the merge of two versions does not depend on the 
common source. The latter is only needed if the two versions 
are to be constructed from the modifications of the common source. 
But once the two spaces 
are known, it is only a matter of deciding which points and bounded-by 
associations they have in common, and where they supplement or possibly 
contradict each other, in order to obtain the topological merge.

As backtracking is the only way to produce existing versions, it follows that 
every version space is a $T_0$-space.
We allow any finite $T_0$-space as 
a possible version space. In particular, 
a version space need not have a unique starting point,  or
even be connected.
This allows to begin with different parts of some spatial model. 
Each  are successively modified, modifications are merged, and in the end 
one unique realisation is obtained through a final merge.
In every step, a valid
topological data type is obtained, and at each merge the occasional
inherent and consistency conflict is resolved whenever it occurs.

To store a DAG  as a topological data type comes natural.
An alternative would be 
a one-dimensional simplicial complex.
However, this increases the size by 
the relation which associates edges with boundary vertices,
plus the additional orientation information of edges.
Hence, increasing the dimension does in fact lead to a decrease in complexity.

\paragraph{Find all versions which were modified in order to obtain a given version.}
This queries for all elements $x$ in the version space 
$V=(\textit{VX},\textit{VR})$ from which a directed path leads to given version $v$. 
These form the \emph{minimal neighbourhood}
\begin{equation}
U_v:=\mathset{x\in \textit{VX}\mid (x,v)\in\textit{VR}^*},
\end{equation}
where $\textit{VR}^*$ is the reflexive and transitive closure of 
$\textit{VR}$. 
This 
set is obtained by 
the following
a simple relational query which for convenience is given in SQL:
\begin{lstlisting}
   select fromv as version
   from poVR
   where poVR.tov = v;
\end{lstlisting}
where \textit{poVR}, again, is the view on $\textit{VR}^{*}$ as described in 
Sec.\ \ref{sec:spadim}.
This minimal neighbourhood yields precisely the version numbers
needed in order to build up the 
space belonging to $v$ from the initial version(s).

\paragraph{Do given versions and their modifications reconstruct a given version?}
The question is whether 
all paths from the initial versions $V_0\subset V$ 
to the given version $v\in V$ 
pass through the given set $W\subseteq V$ 
of 
versions. The answer is given by the following 
consideration:
The minimal neighbourhood $U_A$ of a set $A\subseteq V$
is 
\begin{equation}
U_A:=\mathset{x\in \textit{VX}\mid \exists a\in A\colon
(x,a)\in\textit{VR}^*},
\end{equation}
and is the union of all 
paths ending in $A$.
The 
paths going out of $A$ are given by  
\begin{equation}
\textrm{cl}(A):=\mathset{x\in \textit{VX}\mid \exists a\in A\colon
(a,x)\in\textit{VR}^*}
\end{equation}
which is the \emph{closure} of $A$ in 
$V$.
Hence, the paths from $V_0$ to $v$ are given by 
\begin{equation}
P(V_0,v):=U_v\cap\textrm{cl}(V_0),
\end{equation}
and the paths passing through $W$ are all contained in 
$
P(W):=U_W\cup \textrm{cl}(W),
$
from which we infer that the answer is given by the query:
\begin{equation}\label{query}
\textit{Is it true that $P(v_0,v)\subseteq P(W)$?}
\end{equation}
This breaks down to queries for
minimal neighbourhoods and closures, composed by union, intersection and inclusion queries. 
The relational query 
for the
 minimal neighbourhood of a set $A$ is a slight modification of that for points:
\begin{lstlisting}
   select version from VX
   where version in (
      select fromv
      from poVR
      where tov in (select id from A));
\end{lstlisting}
The closure $\textrm{cl}(A)$ also uses the partial order $\textit{poVR}$, 
and is achieved by reverses the order of pairs, hence  
simply by swapping the attributes 
$\mathit{tov}$ and $\mathit{fromv}$, in the above 
SQL-statement. 
The relational query (\ref{query}) 
can now be easily formulated, but we refrain from writing 
its precise statement.


\section{Levels of Detail}

It is often important to have spatial data at different levels of 
detail.  
In order to enable queries across different levels of detail it is 
necessary to link objects in one LoD 
with their aggregate object in the next 
LoD. 
As all objects in the finer LoD 
have a unique counterpart in the coarser LoD, 
we have a
\emph{generalisation function} $g\colon X\to Y$
from the fine space $X$ to the coarse space $Y$. 
One important consistency rule
is that objects that are ``close'' 
to each other must generalise to objects which are also ``close'' 
to each other. 
This ``closeness'' is a topological notion and  
determined by the bounded-by relation 
defining the topology of a finite space. 
Respecting that relation is nothing but to require that $g$ be 
\emph{continuous} \cite[p.\ 508]{Alex:DS}. 
More precisely, $g$ is a \emph{continuous function} 
between spaces $(X,R)$ and $(Y,S)$ if 
every bounded-by association for $X$ is mapped to a (possibly indirect) bounded-by association in $Y$: either $g(x_1)=g(x_2)$ or 
\begin{eqnarray*}
(x_1,x_2)\in R\;
\Rightarrow 
\exists\, y_1,\dots, y_m \in Y\colon (g(x_1),y_1), \dots,(y_m,g(x_2))\in S
\end{eqnarray*}
This rule is  equivalent to the usual definition of continuous function from topology \cite[\S 4]{McCord}: 
namely that the pre-image of an open set be open \cite[Ch.\ I \S 5]{Jaenich:Topology}.
Notice that the subspace topology from the last paragraph of 
Sec.\ \ref{sec:tempdim} 
is defined in such a way that the inclusion function is continuous.

Another consistency rule is that $g$ be surjective. Then 
every object in the coarse space is indeed the generalisation of an object 
in the fine space.

\begin{figure}[h,ht]
\begin{equation*}
\xymatrix@=20pt{
*\txt{}@{-}\ar@{-}[r]&*\txt{}\ar@{-}[d]^(.65){*}&\\
&    *+[F]{\tt SpatialObject}
\ar`/0pt[r]^(.65){*}`[rd]^{\displaystyle\tt GeneralisesTo
                           \atop
                           {\displaystyle\tt\{surjective,
                           \atop
                           \displaystyle\tt \ continuous\}}} `[d][]^(.65){0..1}
\ar@{-}`l[l]^(.65){*}[ul]^{\displaystyle\tt BoundedBy}
&&\\
&&
}
\end{equation*}
  \caption{The family of surjective and continuous \texttt{GeneralisesTo}-functions and the \texttt{BoundedBy}-relation on objects.}
  \label{fig:lod}
\end{figure}
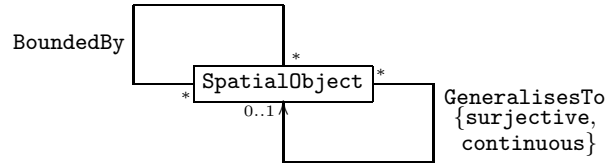

For the classical model 
\begin{equation}\label{classicalmodel}
\textit{Solid}\to\textit{Face}\to\textit{Edge}\to\textit{Vertex}
\end{equation}
a continuous function $g$ for generalisation purposes
implies the explicit modelling of up to 
$16=4^2$ possible types of association pairs, because $g$ can map one class
to any other class, and there is no reason
to forbid the mapping
of one class to a certain other class.
Extending this to spaces of arbitrary dimension $n$ 
(e.g.\ space-time etc.) gives $(n+1)^2$ 
different LoD associations to be modelled explicitly in order to form one single
generalisation function.
So, the classical model (\ref{classicalmodel})
considerably increases the complexity 
for describing functions. 
But if LoD associations are modelled on a common
class of primitive objects, then
the complexity of the class model decreases substantially. 
In the UML diagram of Fig.\ \ref{fig:lod}, 
this class is called \texttt{SpatialObject}. 
Instead of several dimension-dependent bounded-by associations
there is simply one generic \texttt{BoundedBy} association between arbitrary 
objects of a finite space in the most general (topological) sense.  
Functions 
are incorporated 
as a \texttt{GeneralisesTo} association
respecting the consistency rule: 
``continuous  functions that are surjective between two consecutive LoDs''.
Of course, further consistency rules can be imposed 
if necessary.

Assume that it is of interest 
if there is a path from $a$ to $b$
inside a given subset $A$ of a space $X$ with 
a generalisation function $g\colon X\to Y$. 
It would be 
economic to transport the question to $Y$ 
and ask for a path
from $g(a)$ to $g(b)$ inside the set $g(A)$,
and then infer 
back to $X$.
If $A$ is connected, the answer 
is ``yes''.
It is known that if 
$A$ is connected, then 
$g(A)$ is also connected \cite[Ch.\ 7, Thm.\ 1]{Adamson:TW}.
In case the pre-image of every connected subset of $Y$ is
connected,  $g$ is called \emph{monotonic} \cite[Ch.\ V, \S 46]{Kuratowski68}.
For monotonic generalisation functions it follows that if $A$ is the full
pre-image of $g(A)$, then the path query $a\leadsto b$ inside $A$
can be reduced to 
the path query $g(a)\leadsto g(b)$ inside $g(A)$.
In general, one has $A\subseteq g^{-1}(g(A))$, so that 
a positive answer for $g(a)\leadsto g(b)$  yields the existence of
paths $a\leadsto b$ 
inside $g^{-1}(g(A))$ which need to be checked upon leaving $A$ or not.
Of course, by continuity, a negative answer for $g(a)\leadsto g(b)$
inside $g(A)$ implies a negative answer for $a\leadsto b$ inside $A$. 
The upshot is that the consistency rule ``monotonic''
allows to use $g$ for accelerating
the connectivity query for subsets $A$  which are
pre-images of generalised sets by delegating the query to the 
set $g(A)$ which in general is a smaller data set than the original $A$.
Hence, the following \emph{filtering}  approach can be applied:
\begin{algorithm}[Path query]\label{monpathalg}
\emph{Input.} Monotonic generalisation function 
\newline $g\colon X\to Y$, $A\subseteq X$, and $a,b\in A$. 

\smallskip\noindent
\emph{Result.} ``Yes'' if $a\leadsto b$ in $A$, otherwise ``No''.

\smallskip\noindent
\emph{Step 1.} Compute $g(a),g(b),g(A)$. Determine, if $\exists\; g(a)\leadsto g(b)$ inside $g(A)$.

\smallskip\noindent
\emph{Step 2.} If ``No'', then $\textit{answer}=\textit{No}$.
Otherwise, determine if $g^{-1}(g(A))\subseteq A$.

\smallskip\noindent
\emph{Step 3.} If ``Yes'', then $\textit{answer}=\textit{Yes}$.
 Otherwise, determine if $\exists\; a\leadsto b$ inside $A$.

\smallskip\noindent
\emph{Step 4.} If ``Yes'', then $\textit{answer}=\textit{Yes}$.
Otherwise, $\textit{answer}=\textit{No}$.

\smallskip\noindent
\emph{Output.} $\textit{answer}$.
\end{algorithm}

\paragraph{Example.}
\begin{figure}[h]
\begin{equation*}
\xymatrix@=4pt{
&*\txt{}\ar@{-}[ddl]\ar@{-}[rrrr]&&&*\txt{}\ar@{-}[rrrrr]\ar@{-}[ddddd]
&&&&&*\txt{}\ar@{-}[ddddd]
\\
&&&&&&*\txt{}\ar@{-}[dr]&&&&&&&*\txt{}\ar@{-}[rr]&&
*\txt{}\ar@{-}[rrr]*\txt{}\ar@{-}[ddd]&&&*\txt{}\ar@{-}[ddd]
\\
*\txt{}\ar@{-}[dddr]&&A&&&*\txt{}\ar@{-}[ur]&C&*\txt{}\ar@{-}[dl]
&&&\ar[rr]&&&
&A'&&\bullet C'
\\
&&&&&B&*\txt{}\ar@{-}[ul]&&&&&&&&&&B'
\\
&&&&&&&&*\txt{}&&&&&*\txt{}\ar@{-}[uuu]&&*\txt{}&&&*\txt{}\ar@{-}[lllll]
\\
&*\txt{}\ar@{-}[rrr]&&&*\txt{}\ar@{-}[dr]&&&&&*\txt{}
\\
&&&&&*\txt{}\ar@{-}[urrrr]&&
}
\end{equation*}
\caption{A generalisation to a non-classical space.}\label{fig:generalise2}
\end{figure}
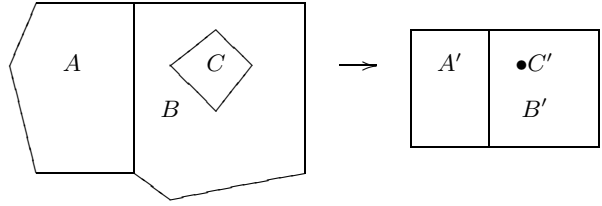
Assume the generalisation of a region subdivided into polygons $A,B,C$,
as depicted in Fig.\ \ref{fig:generalise2}.
There is a monotonic generalisation function to $A',B',C'$.
Now polygon $C$ generalises to 
vertex $C'$
 in the interior of polygon $B'$.
However, the classical model (\ref{classicalmodel}) does not allow
vertex $C'$ to be in a direct bounded-by association with a face.
Consequently, the information that there is a path from anywhere in $A'$
to $C'$ inside the generalised region cannot be inferred from
the classical model
without resorting to the underlying geometry. 
This has grave consequences.
Namely, the geometry and topology of the generalised region
are by force inconsistent: geometry says that $C'$ lies inside
the face $B'$, but the classical model forces $C'$ to be
topologically disconnected from $B'$.
Consequently, the generalisation function is not
 continuous!
And since the statement: ``$B'$ is bounded by $C'$''
cannot be explicitly modelled, 
it has to be inferred from
the position of $C'$. But then, a possible
position error of $C'$ causing this vertex to be outside
$B'$ cannot be corrected by the topological model.
But in that case, the incorrect geometry would be 
consistent with the classical topology model \dots

This is remedied by topological data types.
A direct bounded-by relation $R$ associates $B'$ with 
the four  lines adjacent to $B'$ 
which are themselves, through $R$, 
bounded by 
the four corners of $B'$ 
which are 
not bounded by other objects. Hence, the latter
are vertices, implying 
that $B'$ must be a face. 
And $R$ associates $B'$ with $C'$ which in turn is not bounded by
an object. Hence, $C'$ is a vertex at the boundary of object $B'$.
And now the generalisation function is continuous, and furthermore
monotonic. So, the question whether there is a path from anywhere in $A$ 
to $C$ can be delegated to the simpler generalised region.

We now show how interpolation between different LoDs 
is made possible. Assume that there is a chain $X_1\to\dots\to X_m$
of generalisation functions, 
and each space $X_i$ is embedded in some $\mathbb{R}^n$. 
We assign to each $X_i$ an extra level-coordinate $i\in\mathbb{R}$, leading
to an embedding into $\mathbb{R}^{n+1}=\mathbb{R}^n\times\mathbb{R}$,
and each LoD $i$ sits inside some slice $\mathbb{R}^n\times\mathset{i}$.
The generalisation function $g_i$ associates each $x\in X_i$ with some 
$g_i(x)\in X_{i+1}$. If $x$ and $g_i(x)$ are vertices, 
then they have, by our model,
 coordinates in $\mathbb{R}^{n+1}$, and an interpolating function between 
$x$ and $y$ can be defined.
So, by assigning to \emph{every} element $x$ of every $X_i$ coordinates (e.g.\
by taking a representative in the interior of the geometric realisation of $x$),
it becomes possible to define interpolation functions
between any object $x$ and $g_i(x)$. The interpolation function can be given as
a family of functions $f_x\colon[0,1]\to \mathbb{R}^{n+1}$ with $f_x(0)=x$ and 
$f_x(1)=g_i(x)$,
and as long as $t\in[0,1)$ the objects $f_x(t)$ are considered  a 
placeholder for $x$ inside $X$, but when $t=1$ they  become generalised to
$f_x(1)=g_i(x)$. In this way, the topology of $X_i$ stays unchanged while 
$t\in[0,1)$ and becomes that of $X_{i+1}$ as soon as $t=1$.
The geometry can be made to gradually shrink from one LoD to the next.
By giving coordinates to all elements, 
and not only the vertices, it becomes possible 
in a geometric realisation of 
a continuous zoom, for each vertex to have a unique trajectory,
and so a unique position inside each slice between two LoDs.
This includes also those vertices generalising 
to elements which are not vertices. In fact, 
unique trajectories become possible for all elements through the
positions of their geometric representatives (i.e.\ coordinates) in 
$\mathbb{R}^{n+1}$. 

From 
above, we derive the following relational schema for
generalisations which allow interpolations between LoDs.
Notice that the table $\mathit{Vertex}$ from Sec.\ \ref{sec:spadim}
 is now called $\mathit{Point}$,
as it may contain elements which are not vertices.
\begin{eqnarray*}
&X(\underline{\textit{id}},\underline{\textit{lod}},\textit{gid},\textit{glod},\textit{atts}),
\quad
(\textit{gid},\textit{glod})\stackrel{\textrm{CFK}(R)}{\longrightarrow} X
\\
&R(\underline{\textit{ida}},\underline{\textit{idb}},\underline{\textit{lod}}),\quad
(\textit{ida},\textit{lod})\stackrel{\textrm{FK}}{\longrightarrow}X,\;
(\textit{idb},\textit{lod})\stackrel{\textrm{FK}}{\longrightarrow}X
\\
&\mathit{Point}(\underline{\textit{pid}},\underline{\textit{lod}},x,y,z,t),\quad
(\textit{pid},\textit{lod})\stackrel{\textrm{FK}}{\longrightarrow}X
\end{eqnarray*}
Here, $(\textit{gid},\textit{glod})\stackrel{\textrm{CFK}(R)}{\longrightarrow}X$ denotes a 
\emph{continuous foreign key}, i.e.\ a foreign key which defines a continuous function from the set 
of all tuples in $X$ having $(\textit{gid},\textit{glod})\neq (\texttt{NULL},\texttt{NULL})$ to $X$ 
with respect to the topology generated by $R$. 

The attribute $R.\mathit{lod}$ in both foreign keys from $R$ to $X$ gives a 
disjoint union of an indexed family of spaces \cite[Ch.\ I, \S 3]{Jaenich:Topology}. 
This schema models each LoD in its entirety, 
whereas \cite{Oosterom:VarioScale} propagates the idea 
that objects in one LoD not collapsing with other objects in the next 
LoD need not be repeated in the model. 
It would be interesting to compare both approaches.


\section{Integrating the Different Spaces}


Here, we put together the individual 
schemas for space-time, version and scale
 to one single schema.
The different data models are integrated 
 by collecting  all tables, incorporating the 
attributes,
and providing the 
foreign keys.
As the LoD-attribute is part of the primary key for the 
point set,  all elements (not only vertices) 
can be given  space-time coordinates. 
This leads to the following 
tables:
\begin{align*}
X(\underline{\textit{id}},\underline{\textit{lod}},&
\textit{gid},\textit{glod},\textit{version},\textit{atts}),\;
R(\underline{\textit{ida}},\underline{\textit{idb}},\underline{\textit{lod}}),\;
\textit{Point}(\underline{\textit{pid}},\underline{\textit{lod}},x,y,z,t)
\\
\textit{DelX}(\underline{\textit{id}},\underline{\textit{lod}},&
\textit{version}),\;
\textit{DelR}(\underline{\textit{ida}},\underline{\textit{idb}},\underline{\textit{lod}},\textit{version}),\;
\textit{VX}(\underline{\textit{version}}),\;
\textit{VR}(\underline{\textit{fromv}},\underline{\textit{tov}})
\end{align*}
And the corresponding foreign keys are:
\begin{align*}
X.\textit{version}\stackrel{\textrm{FK}}{\longrightarrow}\textit{VX},&\quad
(X.\textit{gid},X.\textit{glod})\stackrel{\textrm{CFK}(R)}{\longrightarrow}X
\\
(R.\textit{ida},R.\textit{lod})\stackrel{\textrm{FK}}{\longrightarrow}X,&\quad
(R.\textit{idb},R.\textit{lod})\stackrel{\textrm{FK}}{\longrightarrow}X
\\
\end{align*}
\begin{align*}
(\textit{Point}.\textit{pid},\textit{Point}.\textit{lod})\stackrel{\textrm{FK}}{\longrightarrow}X,&\quad
(\textit{DelX}.\textit{id},\textit{DelX}.\textit{lod})\stackrel{\textrm{FK}}{\longrightarrow}X
\\
(\textit{DelR}.\textit{ida},\textit{DelR}.\textit{lod})\stackrel{\textrm{FK}}{\longrightarrow}X,&
\quad
(\textit{DelR}.\textit{idb},\textit{DelR}.\textit{lod})\stackrel{\textrm{FK}}{\longrightarrow}X,
\\
\textit{VR}.\textit{fromv}\stackrel{\textrm{FK}}{\longrightarrow}\textit{VX},&
\quad
\textit{VR}.\textit{tov}\stackrel{\textrm{FK}}{\longrightarrow}\textit{VX}
\end{align*}
All spaces made of polytopes in Euclidean space-time $\mathbb{R}^4$ 
can be consistently modelled with their versionings
and generalisations.
Although the combinatorial dimension 
can be arbitrary,  the element coordinates are  $x,y,z,t$. 
Introducing more coordinates
increases the dimension of the embedding 
space $\mathbb{R}^n$.
Other semantic data can be 
linked to the model by extending the 
  database schema.

\paragraph{In which versions is there a path $a\leadsto b$ inside region $A$?}
This query across versions and LoDs 
is 
answered with the 
schema above.
Assume that $a,b\in X$ and 
each point of  $A\subseteq X$ are given by 
$(\textit{id},\textit{lod})$.
First, find all versions of $A$ containing $a$ and $b$. 
As elements enter $X$ 
in the version given by  \textit{version},
and leave $X$  in the version stored in $\textit{DelX}$, 
the versions 
containing $a$ and $b$ are  the 
intersection of the version intervals for the two points.
These 
intervals are determined by the relation $\textit{VR}$
on the version table $\textit{VX}$. 
For each corresponding \textit{version}-value, there is a different version 
of $A$ as a subspace of $(X,R)$, 
determined from $X,\mathit{DelX},\mathit{DelR}$,
 and for each version containing $a$ and $b$, the query can be answered
as follows. If $a$ and $b$ are in the same LoD, then 
its topology,
stored in $R$, can be used. If $a$ and $b$ are in
different LoDs, then the continuous foreign key $\mathrm{CFK}(R)$
is used for determining paths between those  
LoDs.
If $\mathrm{CFK}(R)$ is monotonic, the query can be accelerated
by 
Algorithm \ref{monpathalg}. 
This query uses the whole schema, except for \textit{Point}.
An explicit formulation in SQL is left to the  reader.


\paragraph{The Integrated 6D+ Space.}
So far, we have a system of models 
for ``elementary'' spaces (cf.\ Sec.\ \ref{sec:dim}). 
To show how these can be combined into one space, 
we first 
extract the LoD space $(\mathit{LoDX},\mathit{LoDR})$ by two queries that 
project $\mathit{Xv}$ and $\mathit{Rv}$, a version $v$ of the stored spaces, onto its two 
\textit{lod} attributes:
\begin{lstlisting}
   create view LoDX as 
      select lod from Xv union select glod as lod from Xv; 
   create view LoDR as select lod,glod from Xv; 
\end{lstlisting}
A query converts this into the 1D edge graph $(V,RV)$, $V$ being the 
``union'' of $\mathit{LoDX}$ with $\mathit{LoDR}$ after duplicating the 
identifiers $x$ in $\mathit{LoDX}$ into pairs $(\mathit{lod},\mathit{glod})=(x,x)$ to make it union compatible with $\mathit{LoDR}$.  
The relation $RV$ contains $((a,b),(a,a))$ and $((a,b),(b,b))$ 
for every pair $(a,b)$ in $\mathit{LoDR}$. 
Adding the graph $\mathit{Gv}$ of the generalisation function 
in $\mathit{Xv}$ to $\mathit{Rv}$, after making them 
union compatible, yields a new 
topology $\mathcal{T}(\mathit{Gv}\cup \mathit{Rv})$. 
By an \emph{equi join} of  spaces $(\mathit{Xv},\mathit{Gv}\cup\mathit{Rv})$ 
and $(V,RV)$ on 
$\mathit{Xv}.\mathit{lod}$ and $V.\mathit{lod}$, 
we get an \emph{equi-join space}, 
or a topological \emph{pullback} \cite[p.\ 406]{Hatcher:AT}, 
of dimension $ \ge 5$. This is also a combinatorial variant of the 
\emph{mapping telescope} \cite[p.\ 312]{Hatcher:AT}.  
However, for each LoD $i$, except the coarsest, 
it contains two redundant copies of 
one space: the space at LoD vertex $(x,x)$ 
and a homeomorphic copy at  LoD edge $(x,g(x))$. 
So, the integrated 5D+ space has redundant information, 
which is not unusual  with table joins. 
One task of database design is to normalise 
by factoring 
redundant tables into smaller tables. 
Similarly, it is also be possible to integrate the whole 
versioning history and the LoDs, 
which gives an even bigger 6D+ space with even more redundancies.
For the above schema this means: 
Whereas it 
is possible to formulate a query that integrates all spaces into one 
huge space, this space will have anomalies and thus may serve as a view for integrated 
space-time-version-view queries. However, it is not suitable as a 
relational schema for storage because of its anomalies. 
In short: We propose the future development of a topological 
relational database design theory extending 
its relational counterpart that has proved so successful in recent years. 


\section{Implementation}

In Sec.\ \ref{sec:tempdim}, 
the subset $\mathit{Xt}$ at a time point $t$
of a set $X$ in space-time was obtained by a relational selection.  
This was converted into a topological subspace. 
In fact, all basic query operators of relational algebra 
\cite{Codd:RMV2} can be 
turned into \emph{topological relational query operators} 
 operating on spaces and returning spaces, 
analog to their relational counterparts, as 
 demonstrated in \cite{BradPaul:RelTop}. Here, we shortly describe the 
prototype  
on \url{pavel.gik.kit.edu},
currently designed as 
a first experimental implementation of the \emph{semantics} of the query operators. 

There are two classes of topological constructions \cite[Ch.\ 3]{Adamson:TW}: 
the initial (or ``induced'') 
spaces and the final (or ``co-induced'') 
spaces. 
 A relational query operator  on some 
input \emph{spaces} operates on their elements and returns a 
\emph{set} $X$. 
Now the result tuples are linked with the spatial entities 
from the input by functions: 
either from  $X$ 
back to the input 
(intersection, selection, or join), giving an \emph{initial space}, 
or the function maps input entities to $X$ 
(like union, projection), giving a \emph{final space}. 

The prototype is programmed in Common Lisp, and 
has its own simplified 
relational algebra in Lisp syntax. A space can be defined by the \texttt{space} 
constructor whose  input is 
a set, the topology-defining relation and the 
two foreign keys.
 Each basic operator \texttt{op} has its spatial counterpart 
\texttt{op-space}, like \texttt{natjoin-space}, 
or \texttt{project-space}, that 
acts on the sets, 
constructs the corresponding (initial or final) 
topology for the result set and returns a space. 
These operators can be arbitrarily nested. 
The experiences gained 
will help to 
produce a \emph{topological relational database 
management system} 
which should provide the topological data modelling presented 
above as builtin feature. 
It would also provide topological consistency rules, and  could be a 
starting point for the discussion of 
topological data modelling rules
towards 
a topological data modelling theory 
extending the current 
relational modelling theory. 


\section{Conclusion}

A relational database schema based on Alexandrov topology, 
that seamlessly integrates 4D space-time data, 
version histories, and different levels of detail (LoD),
is presented. 
Such a topology can always be represented by a directed acyclic graph, and it 
imposes fewer restrictions than the canonical
\textit{Solid}-\textit{Face}-\textit{Edge}-\textit{Vertex}-model 
for spatial data. 
The gained flexibility and 
simplicity alleviates more sophisticated spatial data modelling, 
and endows spatial data with an Alexandrov topology, which
has practical consequences: 
As topology is fundamental it is likely to have more to 
offer for spatial data modelling than is momentarily used. 
A first contribution 
is a precise 
definition of ``spatial data dimension''. 
A topological version space  
allows the recovery of different versions of a spatial model
by using queries based on topological constructions. 
Among the new consistency rules, 
``continuity'' of foreign keys allows consistent modelling in different LoDs,
and ``monotonicity'' allows accelerated path queries.
Also, topological queries across versions and LoDs are enabled. 



\subsection*{Acknowledgements.}
This work is funded by the DFG with grants BR2128/12-2 and BR3513/3-2.

\bibliographystyle{splncs}
\bibliography{paul}

\end{document}